\documentclass[11pt,titlepage]{article}

\usepackage{epsfig}
\usepackage{bm}
\usepackage{hyperref}
\hypersetup{
  colorlinks,
  citecolor = black,
  filecolor = black,
  linkcolor = black,
  urlcolor = black,
  linktocpage
}

\usepackage[usenames]{color}
\newcommand{\eg}{{\it e.g.}\ }

\begin{document}

\begin{titlepage}
\begin{center}
\begin{title} 
\bf{\Large{Transverse Spin Structure of the Nucleon through 
 Target Single Spin Asymmetry in Semi-Inclusive Deep-Inelastic
 $(e,e^\prime \pi^\pm)$ Reaction at Jefferson Lab
}}
\\
\vspace{2cm}

\vspace{2cm}
 H. Gao$^{1}$, L. Gamberg$^{2}$, J.-P. Chen$^{3}$, X. Qian$^{4}$, Y. Qiang$^{3}$, M. Huang$^{1}$, 
A. Afanasev$^{5}$, \\M. Anselmino$^{6}$, H. Avakian$^{3}$, G. Cates$^{7}$, 
 E. Chudakov$^{3}$, E. Cisbani$^{8}$,
C. de Jager$^{3}$, \\
 F. Garibaldi$^{8}$, B.T.~Hu$^{9}$,
X. Jiang$^{10}$,
K. S. Kumar$^{11}$, X.M.~Li$^{12}$, H.J.~Lu$^{13}$, Z.-E. Meziani$^{14}$,\\ B.-Q.~Ma$^{15}$, 
Y.J.~Mao$^{15}$,
J.-C. Peng$^{16}$, A. Prokudin$^3$, 
 M. Schlegel$^{17}$, \\P. Souder$^{18}$, Z.G.~Xiao$^{19}$,
Y.~Ye$^{20}$, L. Zhu$^{5}$ \\
\vspace{0.3cm}
\it
$^{1}$Duke University,Durham, NC 27708, USA \\
$^{2}$Penn State University-Berks, Reading, PA 19610, USA \\
$^{3}$Jefferson Lab, Newport News, VA 23606, USA \\
$^{4}$ Kellogg Radiation Laboratory, California Institute of Technology, Pasadena, CA 91125, U.S.A.\\
$^{5}$Hampton University, Hampton, VA 23668, USA \\
$^{6}$Universita di Torino and INFN, Sezione di Torino, I-10125, Torino, Italy \\
$^{7}$University of Virginia, Charlottesville, VA 22901, USA \\
$^{8}$INFN, Sezione di Roma III, 00146 Roma, Italy \\
$^{9}$Lanzhou University, Lanzhou, P.R. China\\ 
$^{10}$Los Alamos National Laboratory, Los Alamos, NM, USA \\
$^{11}$University of Massachusetts, Amherst, MA 01003, USA \\
$^{12}$China Institute of Atomic Energy, Beijing, P.R. China\\
$^{13}$Huangshan University, Huangshan, P.R.~China\\
$^{14}$Temple University, Philidalphia, PA 19122, USA \\
$^{15}$School of Physics, Peking University, Beijing, P.R. China\\
$^{16}$University of Illinois at Urbana-Champaign, Urbana, IL 61801, USA \\
$^{17}$Institute for Theoretical Physics, Universit\"at T\"ubingen, D-72076 
T\"ubingen, Germany
$^{18}$Syracuse University, Syracuse, NY 13244, USA \\
 $^{19}$ Tsinghua University, Beijing, P.R. China\\
$^{20}$ University of Science and Technology, Hefei, P.R. China\\ 

\end{title}
\end{center}
\end{titlepage}

\begin{abstract}

Jefferson Lab (JLab) 12 GeV energy upgrade provides a golden opportunity to
perform precision studies of the transverse spin and 
transverse-momentum-dependent structure in the
valence quark region for both the proton and the neutron. In this paper, we
focus our discussion on a recently approved experiment on the 
neutron as an example of
the precision studies planned at JLab.  The new experiment will perform
precision measurements of target Single Spin Asymmetries
(SSA) from semi-inclusive electro-production of charged pions from a 40-cm
long transversely
polarized $^3$He target in Deep-Inelastic-Scattering kinematics using 11 and
8.8 GeV electron beams. This new coincidence experiment in Hall A will employ a newly
proposed
solenoid spectrometer (SoLID). The large
acceptance spectrometer
and the high polarized luminosity will provide precise 4-D ($x$, $z$, $P_T$ and $Q^2$) data on the Collins,
Sivers, and pretzelosity asymmetries for the neutron through the azimuthal
angular dependence.
The full 2$\pi$ azimuthal angular coverage in the lab is essential in controlling the
systematic uncertainties.
The results from this experiment, when combined with the proton Collins
asymmetry
measurement and the Collins fragmentation function determined from the 
e$^+$e$^-$ collision data,
will allow for a quark flavor separation in order to achieve a determination of the tensor 
charge of the d quark to a 10\% accuracy.
The extracted Sivers and pretzelosity asymmetries will
provide important information to understand the correlations between the
quark orbital angular momentum and the nucleon spin and 
between the quark spin and nucleon spin.

\end{abstract}

\section{Introduction}
\label{sec:intro} 
Deep inelastic lepton-nucleon scattering (DIS) experiments have 
played a fundamental role in describing the partonic momentum structure
of hadrons.
The unpolarized parton distribution
functions (PDF) have been extracted with excellent precision over a
large range of $x$ and $Q^2$ from DIS, Drell-Yan and other processes 
after several decades of experimental and theoretical efforts. The comparison 
of the structure functions in the large $Q^2$ range with QCD 
evolution equations has provided one of the best tests of QCD. 

When the target and/or beam are polarized the essential properties of 
spin-angular momentum structure of hadrons is probed. 
Three decades of intensive 
experimental and theoretical investigation have resulted in 
a great deal of knowledge on the partonic origin of the nucleon spin structure.
Motivated by the ``spin crisis'' 
from the European Muon Collaboration experiment in the 
1980s~\cite{Ashman:1987hv},  the longitudinal polarized parton distribution functions have been determined
with significantly improved precision over a large region of $x$ and $Q^2$ 
from polarized deep-inelastic (DIS) experiments carried out at CERN, SLAC, DESY in the 
last two decades, and more recently at JLab and at RHIC from polarized 
proton-proton scattering (see~\cite{Filippone:2001ux,Kuhn:2008sy} 
for reviews and compilation 
of references).  In particular, considerable knowledge has been gained 
from  inclusive DIS experiments  on the longitudinal structure -- the $x$-dependence  and 
the helicity distributions -- in terms of the 
unpolarized (denoted $q^a(x)$ or $f_1^a(x)$) and helicity (denoted $\Delta q^a(x)$ or $g_1^a(x)$) 
parton distribution functions for the various flavors (indicated by $a$).

In more recent experimental and theoretical studies, it has become evident that precise knowledge of the transverse structure of partons  
is  essential to unfold the full momentum and spin 
structure of the nucleon. This concerns in particular the investigations of the chiral-odd 
transversely polarized quark distribution function or 
transversity~\cite{Artru:1989zv}  (denoted as $\delta q(x), h_1(x)$ or 
also $\Delta_T q(x)$) which is probed in transverse spin
polarization experiments.
Like the axial charge $\Delta q^a =\int_0^1 dx\ (g_1^a(x) +g_1^{\bar{a}}(x))$,
the tensor charge $\delta q^a =\int_0^1 dx (h_1^a(x)-h^{\bar{a}}(x))$
is a basic property of the nucleon.  The essential role of the transversity distribution function emerges from 
a systematic  extension of the QCD  parton model to include transverse 
momentum and  spin degrees of freedom.  In this context, semi-inclusive deep-inelastic lepton nucleon
scattering (SIDIS) has emerged as an essential tool to probe both 
the longitudinal and transverse momentum and spin structure of the nucleon. 
The azimuthal dependence in the scattering of leptons 
off transversely polarized nucleons 
is explored through the analysis of transverse single spin 
asymmetries (TSSAs). Recent work~\cite{Mulders:1995dh,Boer:1997nt,Ji:2004xq} predicts 
that these observables are 
factorized convolutions of leading-twist 
transverse momentum dependent parton distributions (TMDs) and 
fragmentation functions (FFs). 
These functions provide {\em essential non-perturbative} 
information on the partonic sub-structure of the nucleon; they 
offer a rich understanding of the motion of partons inside the 
nucleon, of the quark orbital properties, and of spin-orbit correlations.
They also provide essential information on multi-parton correlations at leading-twist, allowing us to explore and 
uncover the dynamics of the quark-gluon structure of the nucleon.

At leading twist if we integrate over the transverse momenta 
of quarks, the three quark distribution functions remain: the unpolarized
parton distribution $f_{1}$, the longitudinal polarized parton
distribution $g_1$, and the quark transversity distribution
$h_1$. 
Besides $f_1$, $g_1$ and $h_1$, 
there are five more transverse momentum dependent
distribution functions~\cite{Mulders:1995dh,Boer:1997nt}. Since these TMDs provide the 
description of the parton distributions beyond the collinear
approximation, 
they depend not only on the longitudinal momentum
fraction $x$, but also on the transverse momentum, $k_T$. An
intuitive interpretation of the $k_T$ dependent 
transversity distribution, $h_{1}$, 
is that it gives the probability of finding a transversely polarized parton 
inside a transversely polarized nucleon with certain longitudinal momentum
fraction $x$ and  transverse momentum $k_T$. 

The JLab
12 GeV upgrade provides a unique opportunity to extend our understanding of nucleon spin and momentum structure 
by carrying out multi-dimensional 
precision studies  of longitudinal and
transverse spin and momentum 
degrees of freedom from SIDIS experiments
with
 high luminosity in combination with large acceptance detectors.  Such a program
will  provide the much needed kinematic reach 
to  unfold the generalized momentum and flavor structure of the nucleon.  
In the next section, we summarize 
the essential role that transverse polarization studies play in unfolding this
structure in SIDIS.

\section{Transverse Structure and Semi-Inclusive DIS}
\label{sec:trans}
The transverse spin and momentum  structure 
of the nucleon  was  discussed 
in the context of  polarized proton-proton
scattering experiments
by Hidaka, Monsay and  Sivers~\cite{Hidaka:1978pi} in 1978
and in  Drell-Yan Scattering by Ralston and 
Soper~\cite{Ralston:1979ys} in 1979. Early theoretical analysis
of  the transversity distribution of the nucleon was studied by Artru~\cite{Artru:1989zv}
 and a more thorough  treatment was given  by Jaffe and 
Ji in early 1990s~\cite{Jaffe:1991kp} where they introduced the tensor charge
in the context of the QCD  parton model.
 
The transversity function is a chirally odd quark distribution function,
and the least known among the three leading twist parton distribution
functions.  It describes the net quark transverse
polarization in a transversely polarized nucleon~\cite{Jaffe:1991kp}.
 In the
non-relativistic limit, the transversity distribution  function
$h_{1}(x,Q^2)$ is the same as the 
longitudinal quark polarization distribution function, $g_1 (x,Q^2)$.
Therefore, the transversity distribution function
probes the relativistic nature of the quarks inside the nucleon.

There are several interesting properties 
of  the quark transversity distribution.
First it does not mix with gluons; that is, it 
evolves as a non-singlet distribution~\cite{Barone:1997fh}
and doesn't mix with gluons under evolution and thus  
has valence-like behavior~\cite{Bourrely:1997bx}. 
Secondly in the context of the parton model it satisfies the 
Soffer bound~\cite{Soffer:1994ww}, which is an inequality 
among the three leading twist distributions, 
$|h^q_1| \le {\frac{1}{2}} (f^q_1 + g^q_1)$,
based on unitarity and parity conservation. 
QCD evolution 
of transversity was studied in Ref.~\cite{Vogelsang:1997ak}, 
where it was
shown that Soffer's inequality holds up to next to leading order (NLO)   
QCD corrections. In the past~\cite{Goldstein:1995ek} 
and more recently~\cite{Ralston:pv}, 
studies have been performed that consider the violation of this bound.
Therefore, it is interesting to experimentally test the Soffer's 
inequality as a function of $Q^2$. 
Lastly, the lowest moment of $h^q_1$ is the tensor charge,
which has been calculated
from lattice QCD~\cite{Gockeler:2005cj} and various
models~\cite{He:1994gz,Ma:1998bq,Gamberg:2001qc,Cloet:2007em,Wakamatsu:2007nc,Pasquini:2005dk}. \textcolor{black}{Due to the valence-like nature of 
the transversity distribution, measuring transversity in the high-$x$
region (JLab kinematics) is crucial to determine tensor charge
of quarks. } The experimental determination of the transversity function is
challenging - it is not accessible in polarized inclusive DIS measurements
when neglecting quark masses - 
$h_1$ decouples at leading twist in an expansion of inverse powers
of the hard scale in inclusive deep-inelastic scattering due to the helicity
conserving property of the QCD interactions. 
However, paired with
another hadron in the initial state \eg double polarized Drell-Yan processes
 with
 (two transversity distributions)~\cite{Ralston:1979ys},
or in the final state, \eg semi-inclusive deep-inelastic~\cite{Collins:1992kk} 
scattering (transversity and Collins fragmentation function), 
leading twist $h_1$ can be accessed without suppression by a hard scale. 

The most feasible  way to access the transversity distribution
 function  is via an azimuthal single spin asymmetry, 
 in semi-inclusive deep-inelastic
lepto-production of mesons on a transversely polarized nucleon target, 
$e\, N^\uparrow\rightarrow e\, \pi \, X$.  In this case 
the other chiral-odd partner is the Collins fragmentation
function, $H_1^\perp$~\cite{Collins:1992kk}. 
Schematically with  factorization, this transverse single
spin asymmetry (TSSA) contains $h_1$ and $H_1^\perp$, 
$A_{UT}\sim h_1\otimes H_1^\perp$
($U\equiv$ unpolarized lepton 
beam, $T\equiv$ transversely polarized target)~\cite{Boer:1997nt}.  
The Collins function is one of a class of so-called 
{\em naive time reversal odd} (T-odd) transverse momentum dependent 
fragmentation functions - the existence of which has far reaching consequences for 
factorization theorems and the application of
color gauge invariance to nucleon structure.

\begin{figure}
\centering
\includegraphics[height=0.685\textwidth,width=0.4\textwidth,angle=0]{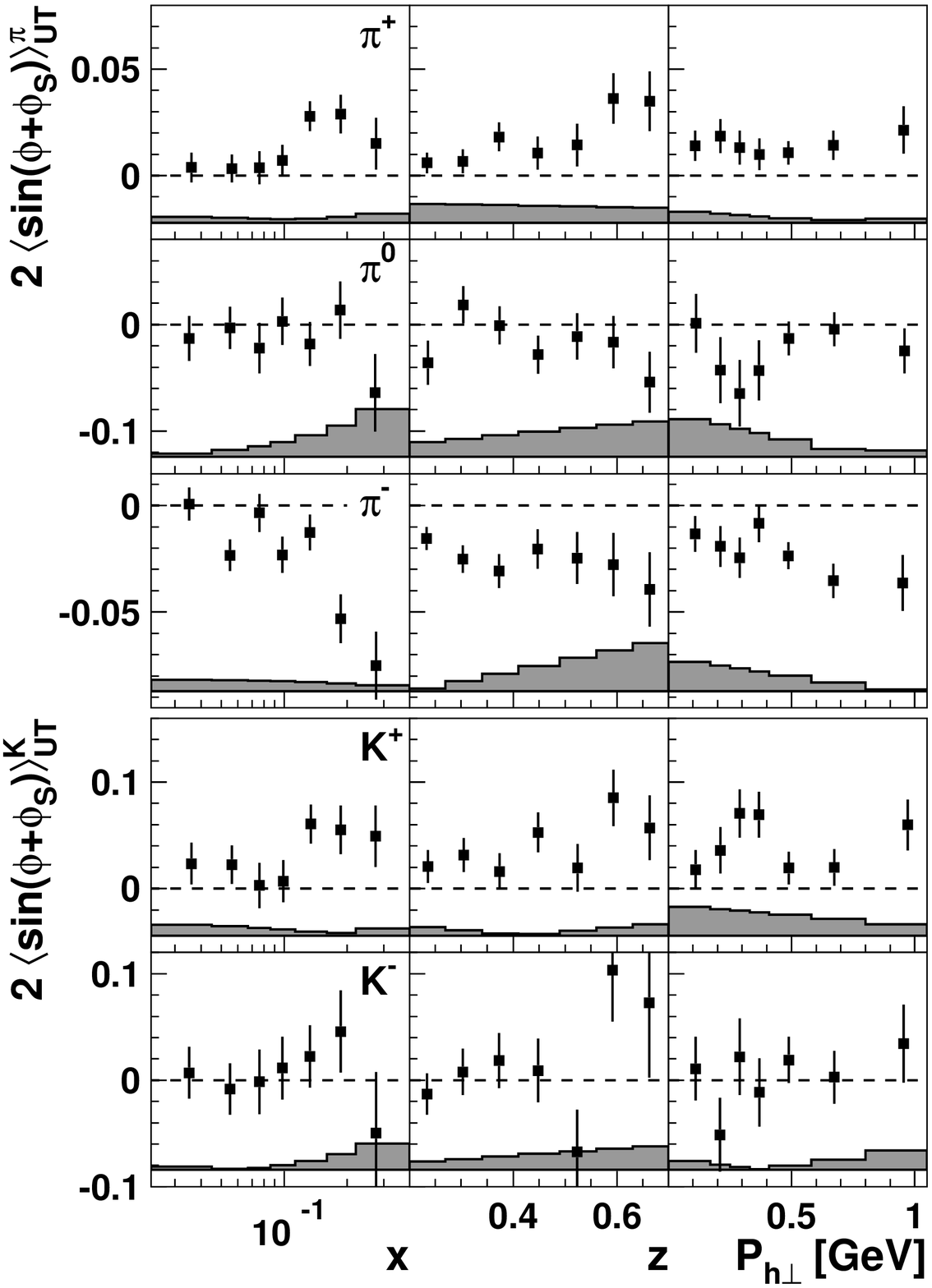}
\hspace{.25cm}
\includegraphics[height=0.685\textwidth,width=0.4\textwidth,angle=0]{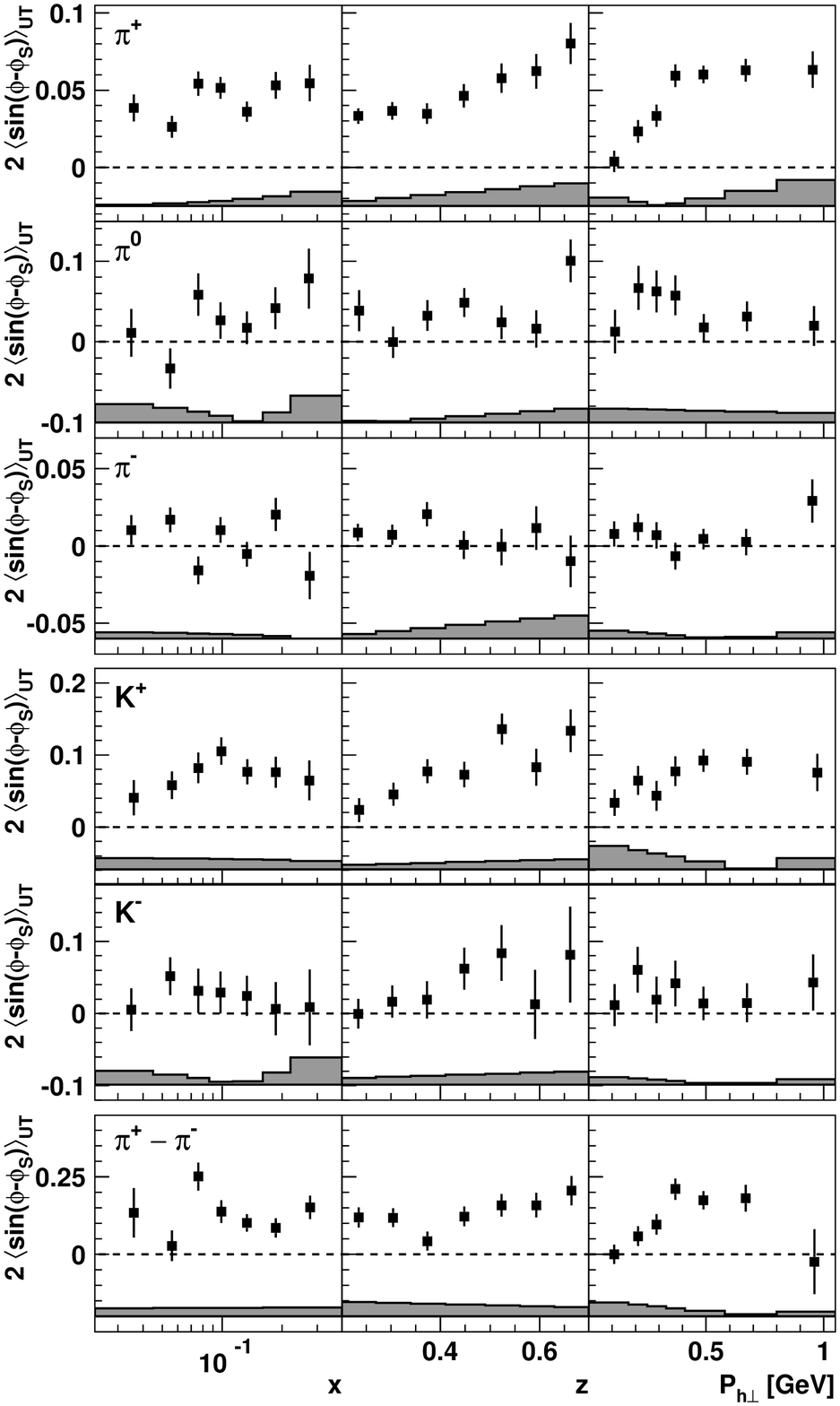}
\caption{
Left Panel: From~\cite{Airapetian:2010ds}, Collins amplitudes for pions and
charged kaons as a function of $x$, $z$, or $p_{\perp}$. The systematic
uncertainty is given as a band at the bottom of each panel. In addition
there is a 7.3\% scale uncertainty from the accuracy in the measurement of
the target polarization.
Right panel:  From~\cite{Airapetian:2009ti}, Sivers amplitudes for pions,
  charged kaons, and the pion-difference asymmetry
  (as denoted in the panels) as functions of $x$, $z$, or
  $p_{\perp}$. The rest of the caption is the same as in the left panel. 
}
\label{hermes_cs}
\end{figure}

\begin{figure}[tb]
\begin{center}
 \includegraphics[width=0.95\textwidth]{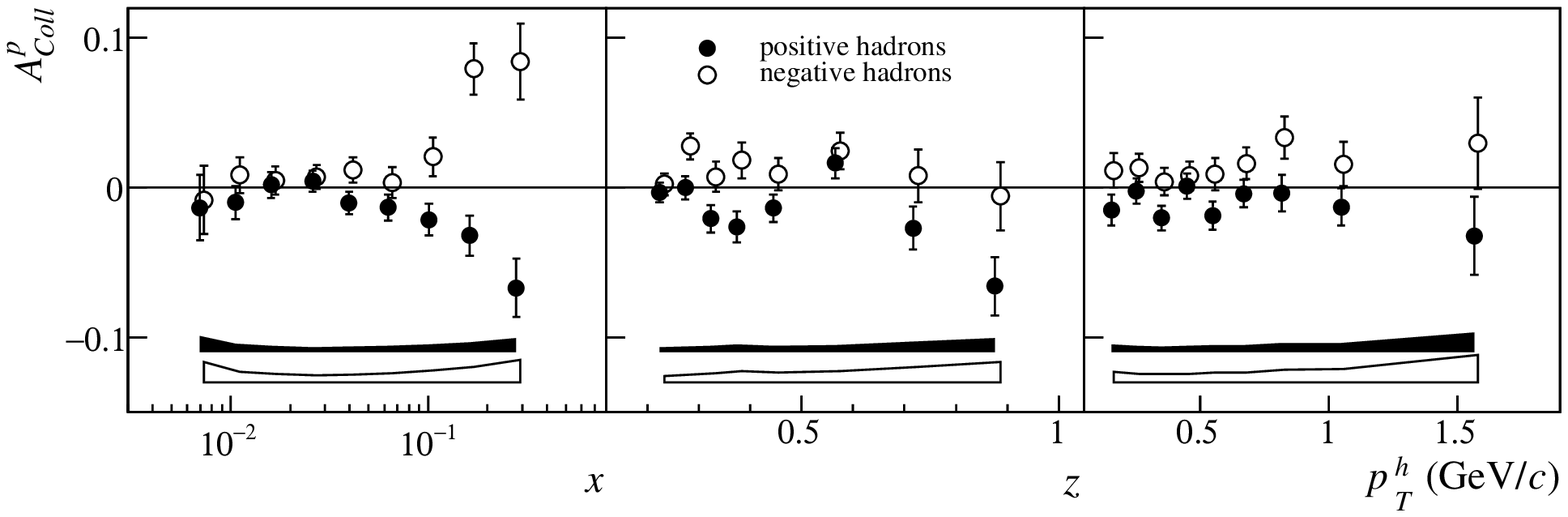}\\
 \includegraphics[width=0.95\textwidth]{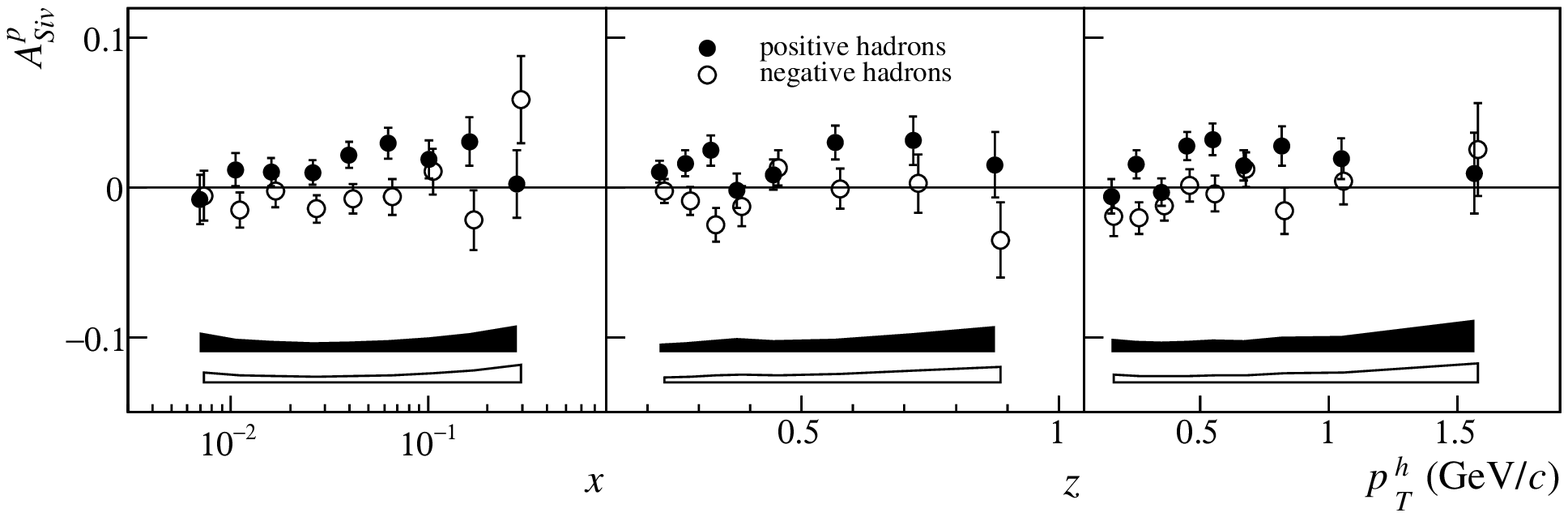}
 \caption{Upper Panel: From~\cite{Alekseev:2010rw}, the 
Collins asymmetry
 as a function of $x$, $z$, and $p_T^h$, for positive (closed points) 
 and negative (open points) hadrons. 
 The bars show the statistical errors.
 The point to point systematic uncertainties have been estimated to be
0.5~$\sigma_{stat}$ for positive and 
0.6~$\sigma_{stat}$ for negative hadrons
and are given by the bands. Lower Panel: From~\cite{Alekseev:2010rw}, the
 Sivers asymmetry
as a function of $x$, $z$, and $p_T^h$, for positive (closed points) 
and negative (open points) hadrons. 
The bars show the statistical errors.
The point to point systematic uncertainties are given by the bands. 
For positive hadrons only, an absolute scale uncertainty of
$\pm 0.01$ has also to be taken into account.
}
\label{compass_cs}  
\end{center}
\end{figure}
The first evidence 
of non-trivial transverse spin effects in SIDIS has
been observed in the transverse single spin asymmetries measured by
the HERMES ~\cite{Airapetian:2010ds,Airapetian:2009ti,Airapetian:2004tw} 
and the COMPASS ~\cite{Alekseev:2010rw} experiments 
where an unpolarized lepton beam is scattered off
a transversely polarized proton target,
$l\, p^\uparrow\rightarrow l^\prime\, h \, X$
(see Figures~\ref{hermes_cs} and~\ref{compass_cs}).
Besides the non-zero Collins asymmetry, which contains $h_{1}$ and $H_1^{\perp}$ discussed previously, another non-zero asymmetry (Sivers asymmetry), was also observed. The Sivers asymmetry is associated with a naive 
T-odd transverse momentum dependent (TMD)
parton distribution function~\cite{Sivers:1989cc}.
By contrast to inclusive deep-inelastic lepton-nucleon scattering 
where transverse momentum is integrated out,
these processes are sensitive to 
 the transverse-momentum scale, $P_T$, which is on the order of the intrinsic quark momentum, $k_T$; that is $P_T\sim k_T$.  
This is evident by considering the generic structure of the TSSA  
for a transversely polarized nucleon target  
which is characterized by interference between helicity flip and helicity
non-flip amplitudes  $ A_{UT}\sim {\rm Im}( f^{*\, +}f^-)$.  In 
the collinear limit of QCD, partonic processes
conserve helicity and Born amplitudes are real~\cite{Kane:1978nd}. 
For this structure to be non-zero at leading twist we must go beyond
the collinear limit where
such a reaction mechanism  requires a recoil  
scale sensitive  to the intrinsic quark transverse momentum. This 
is roughly set by the confinement scale  
$k_T\sim\Lambda_{\rm QCD}$~\cite{Anselmino:1994tv}. Because strongly interacting
processes conserve parity 
transverse spin asymmetries are described by T-odd correlations
between transverse spin $\bm{S}_T$, longitudinal momentum $\bm{P}$ 
and intrinsic quark momentum $\bm{k}_T$~\cite{Sivers:1989cc,Collins:1992kk},
which are depicted by the generic vector product 
$i\bm{S}_T\cdot\left({\bm P}\times {\bm k}_\perp\right)$.
These correlations imply a leading twist reaction mechanism 
which is associated with a naive 
T-odd transverse momentum dependent (TMD)
parton distribution~\cite{Sivers:1989cc} 
and fragmentation~\cite{Collins:1992kk} function
(PDF \& FF). 

A crucial theoretical 
breakthrough~\cite{Collins:2002kn,Belitsky:2002sm,Boer:2003cm}  
was that the reaction 
mechanism is due to non-trivial phases arising  from 
the color gauge invariant property of QCD. 
This leads to the picture 
that TSSAs arise from initial and final state  
interactions~\cite{Brodsky:2002cx,Ji:2002aa,Gamberg:2003ey} (ISI/FSI)
 of the active quark with the soft distribution or fragmentation
remnant in SIDIS, which manifests itself as a gauge link that links the bilocal
quark configuration.
This gauge link gives rise to the final state gluonic interactions 
between the active quark and target remnant.   
Thus, T-odd TMDs are of crucial importance because they
 possess transverse spin polarization structure
as well as the necessary phases to
account for TSSAs at leading twist.   Further work
on factorization theorems for SIDIS indicate that there are two
leading twist T-odd TMDs; the Sivers function, denoted as
$f_{1T}^\perp$  describing the probability density of finding unpolarized partons inside a transversely polarized proton, is one of these functions, while the 
Boer-Mulders function $h_{1}^\perp$  describes the probability density of finding a transversely polarized quark inside a unpolarized nucleon by virtue of the color gauge invariant symmetry of QCD.
All these aforementioned ingredients (TMD, FF, gauge link) enter the 
factorized~\cite{Ji:2004xq} hadronic tensor
for semi-inclusive deep-inelastic scattering.

Exploring the transverse spin structure 
of the TMD PDFs  reveals evidence of a rich
spin-orbit structure of the nucleon.
When the transverse spin and momentum correlations are 
 associated with the nucleon, where the quark remains {\em unpolarized}, 
the Sivers function~\cite{Sivers:1989cc} 
 describes the helicity flip of the nucleon target
in a helicity basis. Since the quark is unpolarized
in the Sivers function, the orbital angular momentum
of the quarks must come into play to conserve overall angular momentum
in the process~\cite{Burkardt:2003yg,Burkardt:2005hp}. 
This result has a far-reaching impact on the QCD theory of generalized angular
momentum~\cite{Brodsky:2006ha}. 
Indeed a partonic description of the Sivers and Boer-Mulders functions 
requires wave function components with nonzero 
orbital angular momentum and thus provides
information about the correlation between the quark orbital angular
momentum (OAM) and the nucleon/quark spin, 
respectively~\cite{Brodsky:2002cx,Brodsky:2006ha}.

Unlike the Sivers and
Boer-Mulders functions, which provide a clean probe of the QCD FSI,
the functions $g_{1T}$ and $h_{1L}^\perp$ are (naive) T-even, and thus do not
require FSI to be nonzero.  Nevertheless,
they also require interference between wave function components that differ
by one unit of OAM and thus require OAM to be nonzero. Finally, the
pretzelosity $h_{1T}^\perp$ requires interference between wave function
components that differ by two units of OAM (e.g. p-p or s-d interference).
Combining the wealth of information from all these functions could be
invaluable for disentangling the spin orbit correlations in the nucleon wave
function, thus providing important information about the quark orbital
angular momentum.

Complementary to Generalized Parton distributions (or Impact
Parameter Dependent distributions), which describe
the probability of finding a parton with certain longitudinal
momentum fraction and at certain transverse position $b$ (1-D momentum
space and 2-D coordinate space), TMDs give a description of the
nucleon structure in 3-D momentum space. Furthermore, by
including the transverse momentum of the quark, the TMDs reveal important
information about the nucleon/parton spin-orbital angular momentum
correlations. 

\section{The Phenomenology TSSAs and TMDs}
\label{sec:phen}

\begin{figure}[t]
\begin{center}
\includegraphics[width=0.55\textwidth,bb= 10 120 500 660,angle=-90]
{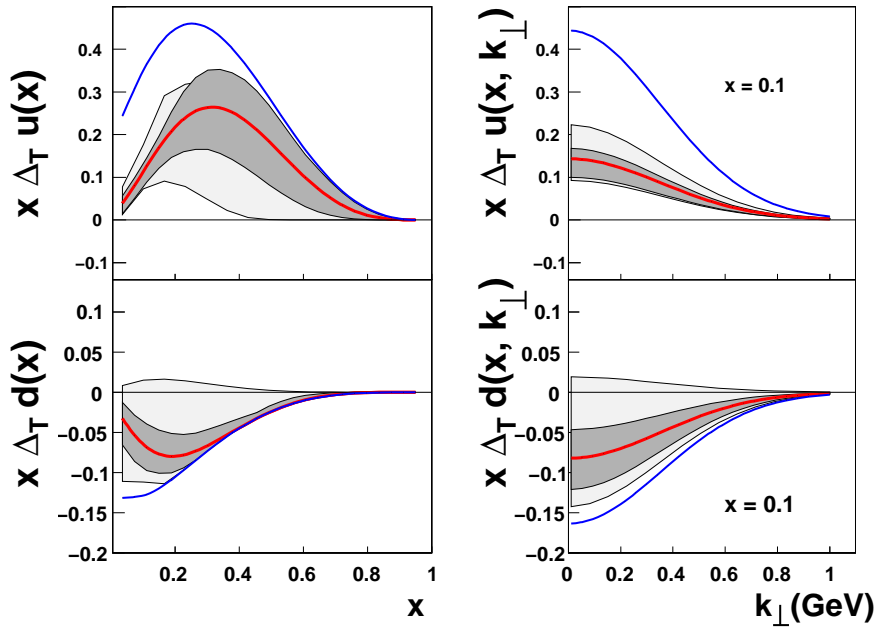}\hskip 2cm
\end{center}
\vskip 0.5cm
\caption{\label{fig:transv}From~\cite{Anselmino:2008jk}, 
the transversity distribution functions
for $u$ and $d$ flavors as determined by a  global fit analysis, at $Q^2$ =
2.4 GeV$^2$~\cite{Anselmino:2008jk}. The Soffer bound~\cite{Soffer:1994ww}
(highest or lowest lines) and the (wider) uncertainty bands of a
previous extraction~\cite{Anselmino:2007fs} are displayed.}
\end{figure}

\begin{figure}[t]
\begin{center}
\includegraphics[width=0.60\textwidth,bb= 10 140 580 660,angle=-90]
{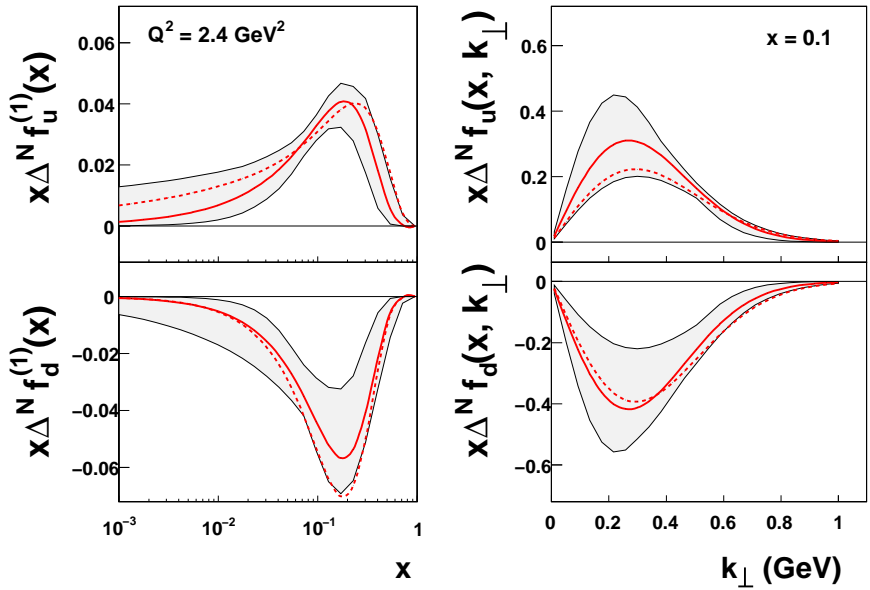}
\caption{\label{fig:sivers-como}
From~\cite{Anselmino:2008sga}, the Sivers distribution functions for $u$ and $d$ flavors, at the scale
$Q^2=2.4$ (GeV$/c)^2$, as determined
by an updated model-dependent fit (solid lines) are compared with those of 
 an earlier fit~\cite{Anselmino:2005ea} of SIDIS data (dashed lines), where $\pi^0$ and kaon
productions were not considered and only valence quark contributions were taken
into account. This plot shows that the Sivers functions previously
found are consistent, within the statistical uncertainty bands, with the Sivers
functions obtained in~\cite{Anselmino:2008sga}.}
\end{center}
\end{figure}

All eight leading twist TMDs can be accessed in 
SIDIS. In this paper, we focus on three of them: transversity, Sivers, and pretzelosity, which  
can be accessed through a transversely polarized target.
There are three mechanisms which can lead to the single (transversely
polarized target) spin azimuthal asymmetries, which are the Collins asymmetry,
the Sivers asymmetry, and the pretzelosity asymmetry. 
As mentioned in Section~\ref{sec:trans} 
the quark transversity function in combination with the chiral-odd
Collins fragmentation function~\cite{Collins:1992kk} gives rise to an azimuthal 
(Collins) asymmetry in $\sin(\phi_h+\phi_S)$, where  
azimuthal angles of both the hadron (pion) ($\phi_h$) and the 
target spin ($\phi_S$) are with respect to the virtual
photon axis and relative to the lepton scattering plane.
The Sivers asymmetry~\cite{Meng:1988cr,Sivers:1989cc,Anselmino:1999pw} 
refers to the azimuthal asymmetry in $\sin(\phi_h-\phi_S)$
due to the correlation between the transverse target polarization of 
the nucleon and the transverse momentum of the quarks, which involves the
orbital angular momentum of the unpolarized quarks~\cite{Brodsky:2002cx,Burkardt:2003yg}.  
The pretzelosity asymmetry is similar to Collins asymmetry 
except it is due to quarks polarized perpendicularly to
the nucleon spin direction in the transverse plane in a transversely
polarized nucleon. It has an azimuthal angular dependence of
$\sin(3\phi_h - \phi_S)$. One can disentangle these angular
distributions by taking the azimuthal moments of the asymmetries as
has been done by the HERMES Collaboration~\cite{Airapetian:2004tw} and the
COMPASS Collaboration~\cite{Alexakhin:2005iw}. 

In recent years a great deal of understanding of transverse spin effects, final state interactions, and the 
spin orbit structure  of partonic-hadronic 
interactions has been gained from model calculations
of the TMDs and fragmentation functions.
%
In particular the final state interactions in TSSAs through 
the Sivers function has been studied 
in spectator models and the light-cone wave function approach~\cite{Brodsky:2002cx,Ji:2002aa,Gamberg:2003ey,Bacchetta:2003rz,Lu:2004au,Gamberg:2007gb,Bacchetta:2008af,Pasquini:2010yu} as well as the bag model~\cite{Yuan:2003wk}.  The Collins function has been calculated 
in~\cite{Amrath:2005gv,Bacchetta:2007wc} while studies of the universality of T-odd
fragmentation functions have been carried out in~\cite{Metz:2002iz,Collins:2004nx,Gamberg:2008yt}.  The Boer-Mulders function has been calculated 
in~\cite{Goldstein:2002vv,Gamberg:2003ey,Lu:2004hu,Gamberg:2007gb,Pasquini:2010yu} and the
spin orbit effects of the pretzelosity function have been studied
in both light-cone constituent quarks models~\cite{Avakian:2008ha,Ma:2009bq,Pasquini:2008ba,Boffi:2009sh}, while model predictions of azimuthal and transverse spin
asymmetries have been predicted in~\cite{Barone:2005kt,Gamberg:2007gb,Barone:2008tn}.

The first model dependent extractions of the 
transversity distribution
have been carried out~\cite{Anselmino:2007fs} 
by combining SIDIS~\cite{Airapetian:2004tw,Airapetian:2009ti,Ageev:2006da,Alekseev:2008dn} data with
 $\rm e^+ e^-$ data~\cite{Abe:2005zx} on the Collins function 
(see Figure~\ref{fig:transv}). Within the uncertainties, 
the Soffer bound is respected. In addition, the extraction of the Sivers 
function~\cite{Anselmino:2008sga,Anselmino:2005ea,Anselmino:2005an,Anselmino:2005nn,Collins:2005ie} 
has been performed by combining
SIDIS data from the HERMES~\cite{Airapetian:2004tw} on the proton and 
COMPASS data~\cite{Alexakhin:2005iw} on the deuteron.
The extracted Sivers functions from~\cite{Anselmino:2008sga} are shown in Fig. 4.
Complementing the  data from the HERMES~\cite{Airapetian:2004tw,Airapetian:2009ti}, COMPASS~\cite{Alekseev:2008dn},
 and BELLE~\cite{Abe:2005zx} experiments, 
 the  recent release from the Jefferson Lab  Hall 
A experiment E06-010~\cite{Gao:2006pr} on the 
neutron (with polarized $^3$He) 
will facilitate a flavor decomposition 
 of the transversity distribution 
function, $h_1$~\cite{Jaffe:1991kp,Barone:2001sp}
and the Sivers distribution function 
$f_{1T}^{\perp}$~\cite{Sivers:1989cc} 
in the overlapping kinematic 
regime.  However 
a model-independent determination of these leading twist functions 
requires data in a wider kinematic range with high precision in {\em four dimensions} 
($Q^{2}, x, z, \bm{P}_T$).

In this paper, we discuss a newly approved experiment~\cite{Chen:2010pr} with an 11-Gev 
electron beam, a high-pressure polarized $^3$He target (as an effective polarized neutron target), and a solenoid 
detection system that will provide data with very high statistical and  excellent systematic precision
over a $Q^2$ range 
of 1 - 4 (GeV/c)$^2$ and a large range of $x$, $z$, and $P_T$ values.
 The determination of the Collins, Sivers
and pretzelosity asymmetries with high precision is very important to test
theoretical predictions of TMDs and to improve our understanding of QCD.

\section{The Experiment}
\label{sec:setup}

\subsection{Overview}

This new experiment consists of a superconducting solenoid magnet, a detector system consisting of forward-angle detectors and large-angle detectors, and a high-pressure polarized $^3$He target positioned upstream of the magnet.
The polarized $^3$He target is based on the technique of spin-exchange optical pumping of hybrid Rb-K alkali atoms.
Such a target was used successfully in the recently completed SSA experiment~\cite{Gao:2006pr} with a 
6-GeV electron beam at JLab and an in-beam polarization of 60-65\% was achieved.
Design studies have been performed with both options of using the Babar and the CDF magnets.
In this paper, we present the design with the option of the CDF magnet with a new yoke added.
The upstream endcap plate will keep the magnetic field and its gradients under control in the target region.
The field strength is simulated using the two-dimensional Poisson Superfish code.
In this design, the absolute magnetic field strength in the target region is about a few Gauss with field gradients $<$~50~mG/cm.
Correction coils around the target will further reduce field gradients to the desired level of $\sim$~30~mG/cm.

\begin{figure}[tbp]
\begin{center}
\resizebox{\textwidth}{!}{\includegraphics{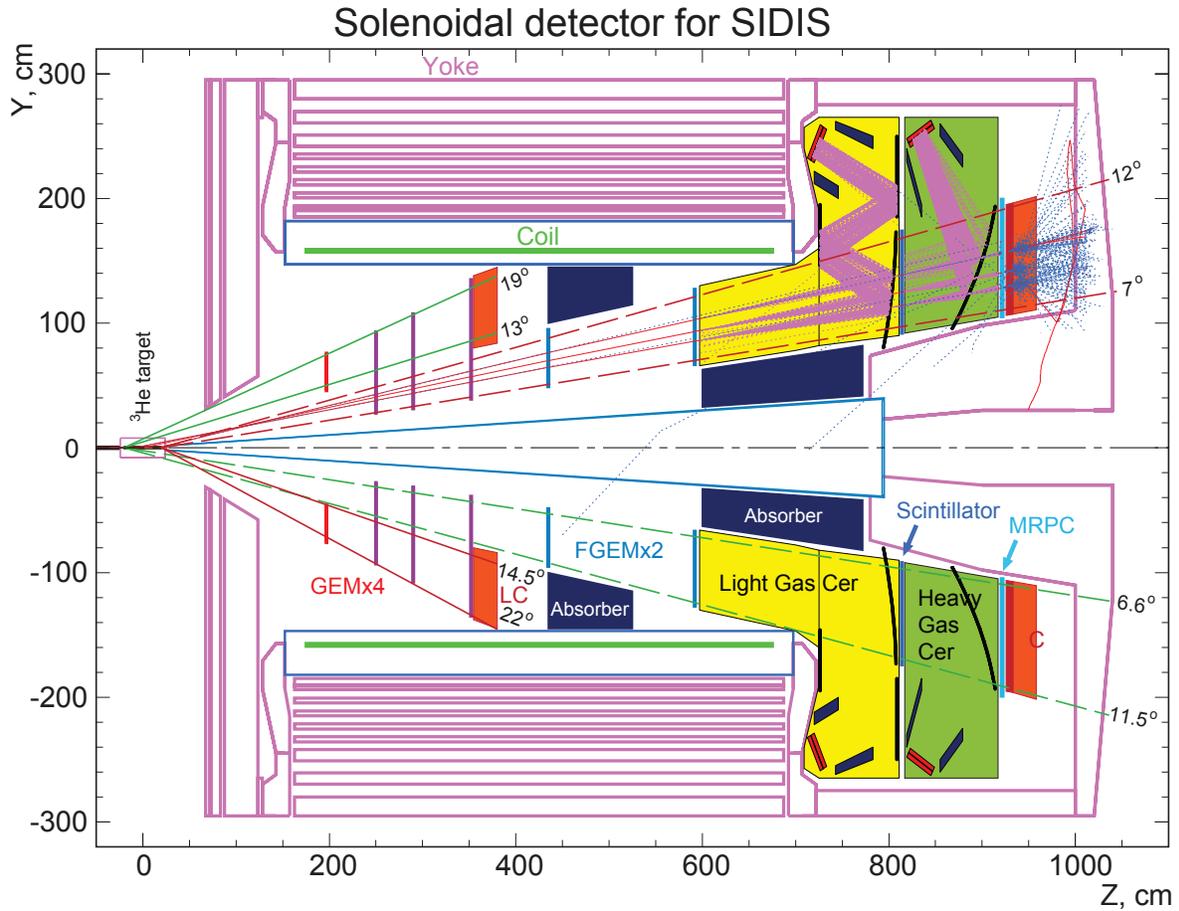}}
\caption{The experimental layout of the SoLID based
  on the option of using the CDF magnet. At forward angles, there are five layers of GEM
  detectors (The first three, in purple color, are shared with the
  large-angle detectors. The other two layers are in blue color.) inside the
  coils upstream of a 2.1-meter long light gas \v{C}erenkov (yellow). 
  One layer of scintillator (dark blue) will
  be placed after the light gas \v{C}erenkov. 
  A 1-meter long heavy gas \v{C}erenkov (in green) is placed after the
  scintillator.
 One layer of Multi-gap Resistive Plate Chamber (MRPC) (light blue) 
  is placed after the heavy gas \v{C}erenkov.
  The electromagnetic calorimeters are shown in orange.}
\label{fig:cdf_layout}
\end{center}
\end{figure}

The layout of the experiment is shown in Fig.~\ref{fig:cdf_layout}.
The acceptance is divided into large-angle and forward-angle regions.
The forward angle detectors cover the polar angle from 6.6 to 12 degrees with a momentum coverage from 0.9 to 7.0 GeV/c while the large angle side covers 14.5 to 22 degrees and 3.5 to 6.0 GeV/c.
The total solid angle is about 57~msr for the forward-angle region and 231~msr for the large-angle region.
Six layers of GEM detectors will be placed inside the coils as tracking detectors.
A combination of an electromagnetic calorimeter, gas \v{C}erenkov counters, a layer of Multi-gap Resistive Plate Chamber (MRPC) and a thin layer of scintillator will be used for particle identification in the forward-angle region.
As only electrons will be identified in the large-angle region, 
a shashlyk-type~\cite{Atoian:2004nm,Atoian:2008nm} electromagnetic calorimeter will be sufficient to provide the pion rejection.

\subsection{The Detectors}

\subsubsection{Tracking and Resolutions}
A total of six layers of GEM tracking detectors will be placed inside the magnet to determine the momentum, angle and vertex of the particles detected.
Five layers of GEM detectors will be used for the forward-angle region.
The first four layers will be sufficient for the large-angle region because the background level 
is expected to be much lower. The GEMs have worked for the COMPASS experiment~\cite{Altunbas:2002nm,abbon:2007nm} in a flux of 30~kHz/mm$^2$, which is much higher than the
estimated rates in our configuration.

To optimize the GEM configuration, a tracking MC simulation was performed, 
and a new tracking algorithm was developed based on the idea of progressive search\cite{Mankel:2004rp}.  The tracking MC study demonstrated that such a GEM configuration will work 
sufficiently well for the planned luminosity~\cite{Chen:2010pr}.

\begin{figure}
\begin{center}
\resizebox{100mm}{100mm}{
\epsfig{figure=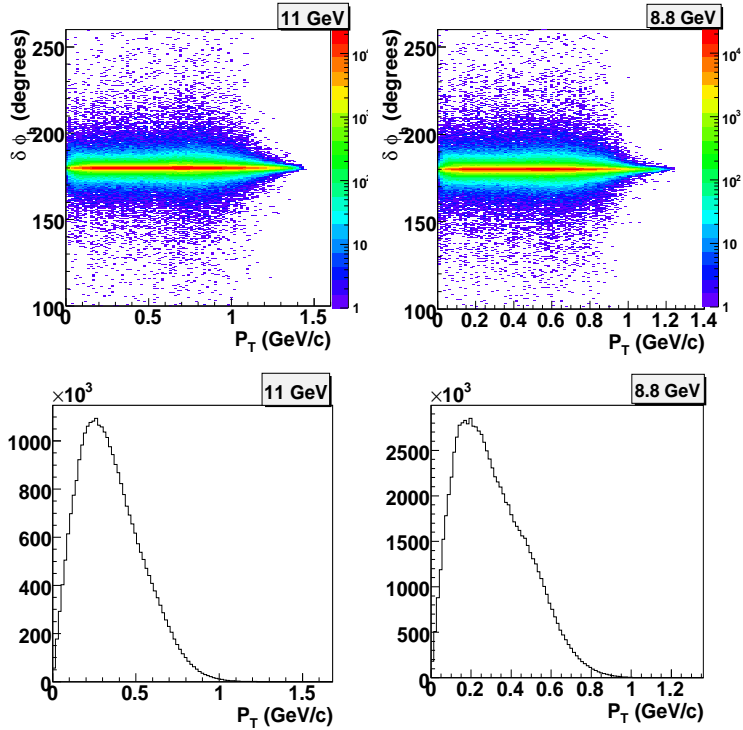}}
\caption{The $\delta \phi_h$ vs $P_T$ are shown in top left (
  right) panel for 11 (8.8) GeV. The $P_T$ coverage are shown 
in bottom left (right)
panel for 11 (8.8) GeV.}
\label{fig:phi_res}
\end{center}
\end{figure}

A simulation was performed to study the reconstruction resolutions with the CDF magnet. With the
 expected GEM resolution of 200 $\mu$m, the average momentum resolution, $\delta p/p$, is about 1.2\%, 
and polar angle resolution is around 0.3 mrad while the azimuthal angular resolution is around 6 mrad. 
The average vertex resolution is about 0.8 cm over the entire momentum range.     
In addition to the polar and azimuthal angular resolution in the lab
frame, the $\phi_{h}$ resolution was also studied and plotted versus the hadron transverse momentum 
$P_{T}$ together with the $P_{T}$ coverage as 
 shown in Fig.~\ref{fig:phi_res}. Here $\delta \phi_{h}$ is
defined as the angle difference between the $\phi_h$ with and without the detector
resolution effects. In Fig.~\ref{fig:phi_res}, no clear resolution
degradation is observed in $\delta \phi_{h}$.

\subsubsection{Electron Identification}
Two sets of electromagnetic calorimeters will be used to identify electrons in the forward and large-angle regions by measuring the energy deposition in the calorimeter through the electromagnetic shower.
As discussed in more details in~\cite{Souder:2009pr}, a ``shashlyk"-type calorimeter can be used inside the magnetic field and it is also radiation resistant.  With a pre-shower/shower splitting, a pion rejection factor of 200:1 can be achieved at $E~>~3.5$~GeV and over 100:1 at $E~>~1.0$~GeV.

To further improve the electron identification in the forward-angle region, two gas \v{C}erenkov detectors will be used. 
Filled with CO$_2$ at 1 atmospheric pressure (n=1.00045), the light gas \v{C}erenkov has a pion momentum threshold of 4.7~GeV/c.
The 2-meter long setup is expected to produce about 17 photoelectrons for high energy electrons.
As the overall background is estimated to be 40~MHz in such a detector, with 30 sectors and a 20-ns coincidence window, a 40:1 pion rejection can be achieved on-line, and the off-line pion rejection can be expected to be better than 80:1.
The 80-cm long heavy gas \v{C}erenkov detector filled with C$_4$F$_{10}$ at 1.5 atmospheric pressure (n = 1.0021) will provide additional suppression for pions with momentum up to 2.2~GeV/c.
With 60~MHz background and about 25 photoelectrons for electrons in the detector, the pion rejection is expected to be better than 50:1.

The E06-010 (6~GeV transversity)~\cite{Gao:2006pr} analysis shows that by requiring coincidence between pions and electrons in the DIS region, the pion contamination can be further reduced by about a factor of 5.
Assuming a similar suppression factor in the kinematics of this experiment,  the overall pion contamination in the electron sample can be controlled to be less than 0.2\% for the coincident SIDIS events.

\subsubsection{Pion Identification}
For the forward-angle region, the identification of $\pi^{\pm}$ with momentum between 0.9 to 7.0 GeV/$c$ will be one of the major goals of the SIDIS experiment.
The CO$_2$ gas \v{C}erenkov and heavy gas \v{C}erenkov detectors will separate pion from heavier hadrons with momentum range of  $4.7 - 16$~GeV/c and $2.2 - 7.6$~GeV/c, respectively.
The background rejections are about 80:1 and 50:1 from the two detectors.

In order to identify low-momentum pions, a multi-resistive plate chamber (MRPC) detector will be inserted after the two \v{C}herenkov detectors and before the forward-angle calorimeter.
A MRPC has been recently used in the STAR detector at RHIC and ALICE at LHC as their Time-Of-Flight detectors and its typical timing resolution is better than 80~ps.
A MRPC does not need PMTs for readout so it can work inside a magnetic field.
The simulated background rate on MRPC is shown to be less than 0.1 kHz/mm$^2$.
Study~\cite{Wang:2009ab} shows that the MRPC can work in an environment of background rate of 0.28 kHz/mm$^2$.
The total path length is around 9 meters from the target and the flight time is calculated by comparing the TOF signal to the beam RF signal.
With a TOF resolution of 100~ps, we can identify charged pions from charged kaons at a minimum rejection factor of 20:1 for a momentum range up to 2.5~GeV/$c$.

\subsection{The Kinematics and Projections}
\label{sec:kine}

\begin{figure}[htbp] 
\centerline{\includegraphics[width=100mm]{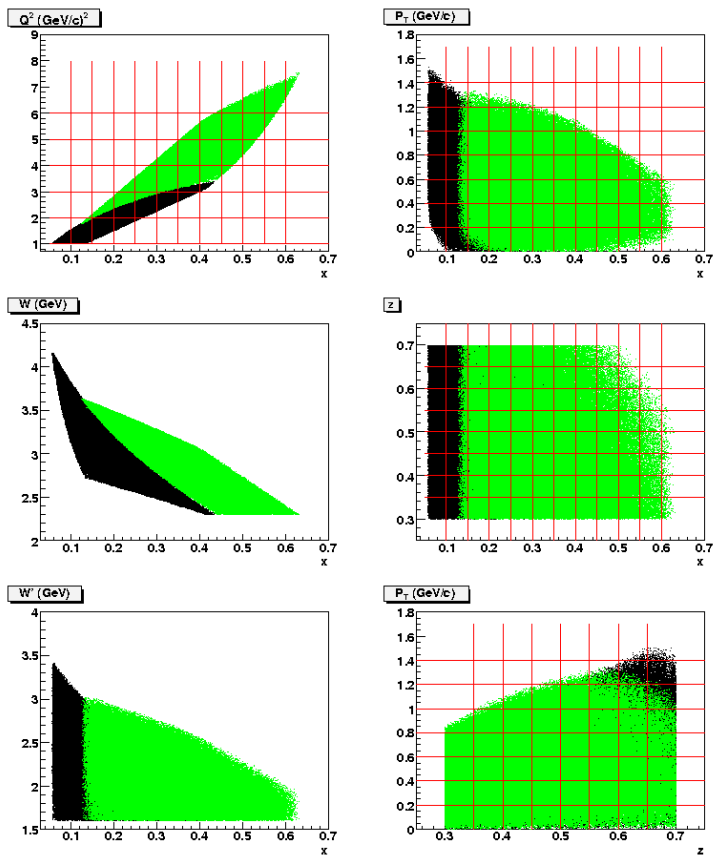}} 
\caption{\label{fig:cdf_kine1} \textcolor{black}{Kinematic coverage for the solenoid
  detector with a 11 GeV electron beam. The black points show the
  coverage for the forward-angle detectors and the green points show the
  coverage for the large-angle detectors.}}
\end{figure}

The phase space coverage is obtained from a detailed Monte Carlo
simulation based on GEANT3 which includes realistic spectrometer
models as well as the target and detector geometry. The coverages in the
$(Q^2,x)$, $(W,x)$ ,$(W^\prime, x)$, $(p_T, x)$, $(z,x)$ and $(p_T,z)$ 
are shown in Fig.~\ref{fig:cdf_kine1} for a 11-GeV incident beam. 
The polar angles of electrons $\theta_e$ and pions $\theta_h$
have coverage from 6.6$^\circ$ to 22$^\circ$ and 6.6$^\circ$ to
12$^\circ$, respectively. The momentum coverages for 
electrons and pions are from 1.0 GeV/c to 7.0 GeV/c. To ensure DIS
kinematics, we will apply cuts for $Q^2 > 1$ (GeV/c)$^2$, W $>$ 2.3 GeV
and W' $>$ 1.6 GeV (missing mass) to avoid the resonance region. 
The final kinematic coverage is $x$ = 0.05-0.65, $Q^2$ = 1.0-8.0 (GeV/c)$^2$,
which covers most of the region where the 
 transversity distributions are significant.    
We choose to detect leading pions with $0.3<z<0.7$ to favor the 
current fragmentation region. 
The resulting transverse momentum of 
the hadron, $P_T$ is between 0 and 1.6 GeV/c for an 11-GeV
incident beam and
between 0 and 1.2 GeV/c for an 8.8-GeV beam. 
 Combining the 11 GeV and
8.8 GeV simulations results together, the total number of bins is
about 1400. This will allow for a map of the SSAs in four dimensions ($x$, $Q^2$, $z$ and $P_T$). 

Due to the nature of the large
acceptance solenoid detector, we will have a complete 2$\pi$ coverage
for $\phi_S$. The full 
$2\pi$ azimuthal angular coverage and a large $\phi_h$ azimuthal
angular coverage are very important in disentangling different
asymmetries (Collins, Sivers and Pretzelosity) to high precisions so that
potential contributions from
other azimuthal angular dependent terms due to higher-twist contributions
($\sin(\phi_S)$ and $\sin(2\phi_h - \phi_S)$) can be separated out.

\begin{figure}[htbp] 
\centerline{\includegraphics[width=150mm,height=90mm]{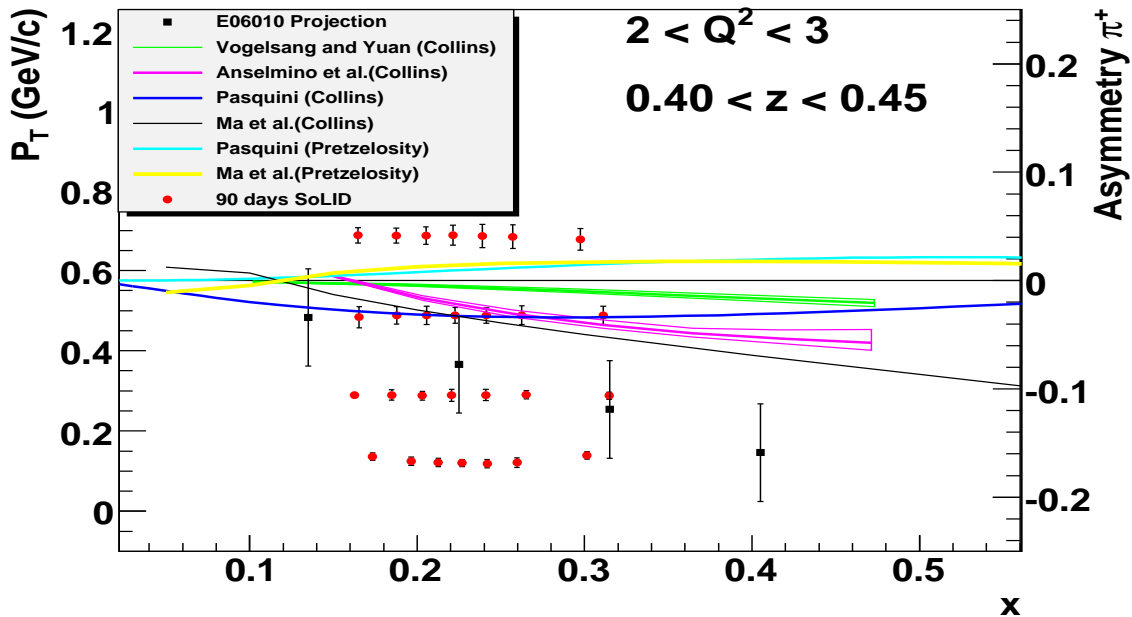}} 
\caption{\label{fig:cdf_proj1} 12 GeV Projections with SoLID. $\pi^+$ 
  Collins/pretzelosity asymmetries at 0.45 $>$ $z$ $>$ 0.4, 3 $>$ $Q^2$ $>$ 2.}
\end{figure}

\begin{figure}[htbp] 
\centerline{\includegraphics[width=150mm,height=90mm]{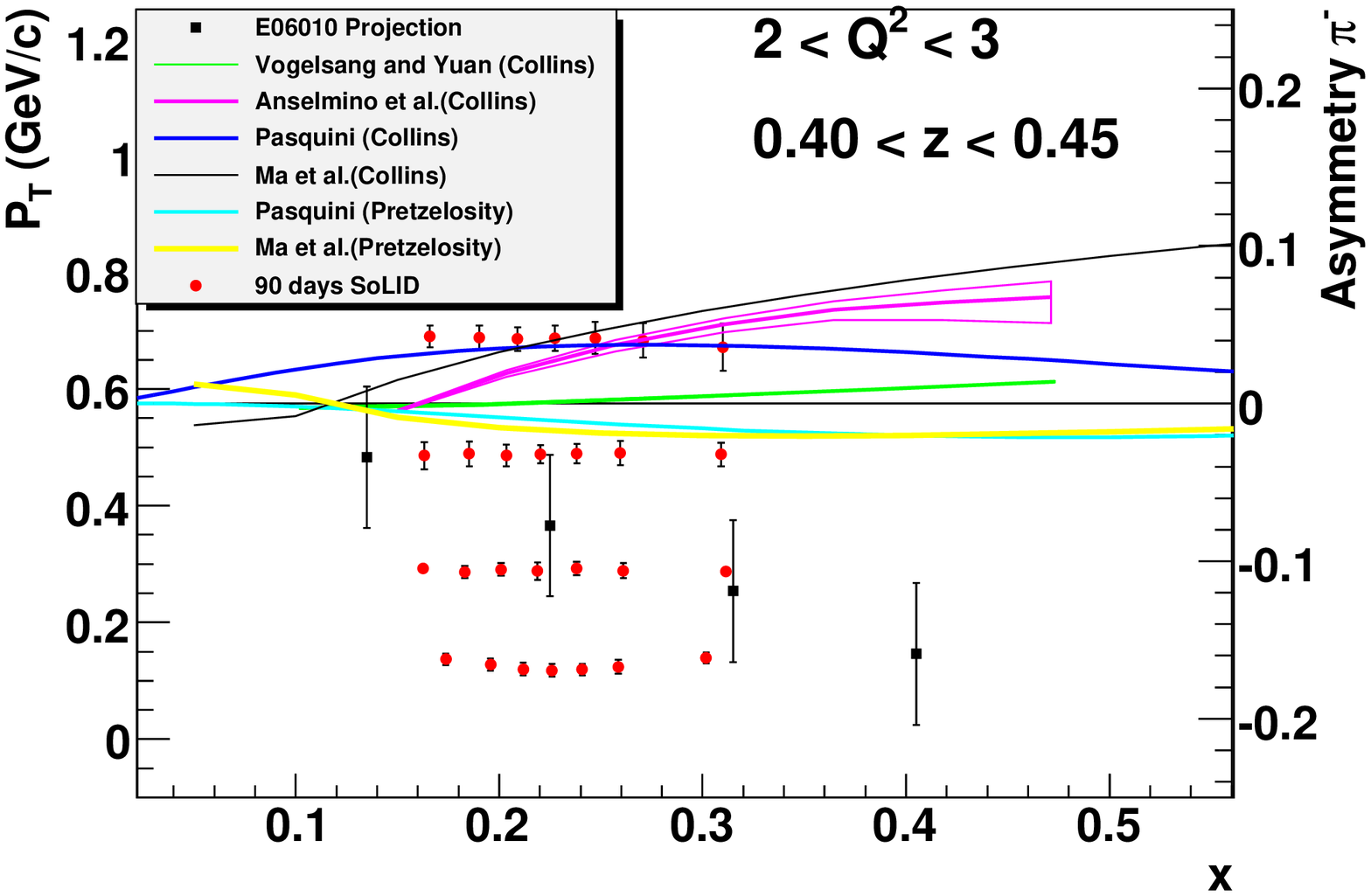}} 
\caption{\label{fig:cdf_proj2} 12 GeV Projections with SoLID. $\pi^-$
  Collins/pretzelosity asymmetries at 0.45 $>$ $z$ $>$ 0.4, 3 $>$ $Q^2$ $>$ 2.}
\end{figure}

\begin{figure}[htbp] 
\centerline{\includegraphics[width=150mm,height=90mm]{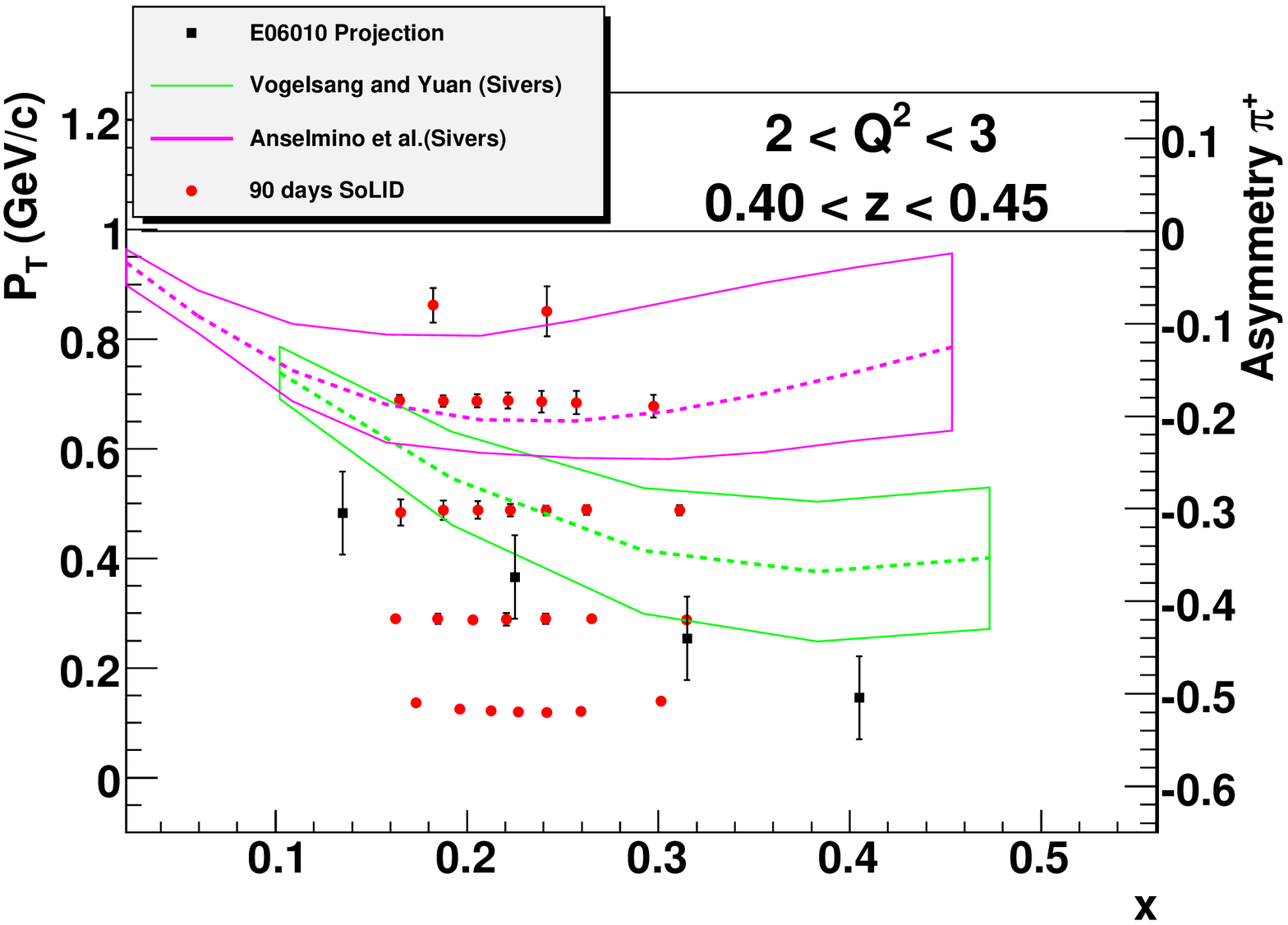}} 
\caption{\label{fig:cdf_proj3} 12 GeV Projections with SoLID. $\pi^+$
  Sivers asymmetries at 0.45 $>$ $z$ $>$ 0.4, 3 $>$ $Q^2$ $>$ 2.}
\end{figure}

\begin{figure}[htbp] 
\centerline{\includegraphics[width=150mm,height=90mm]{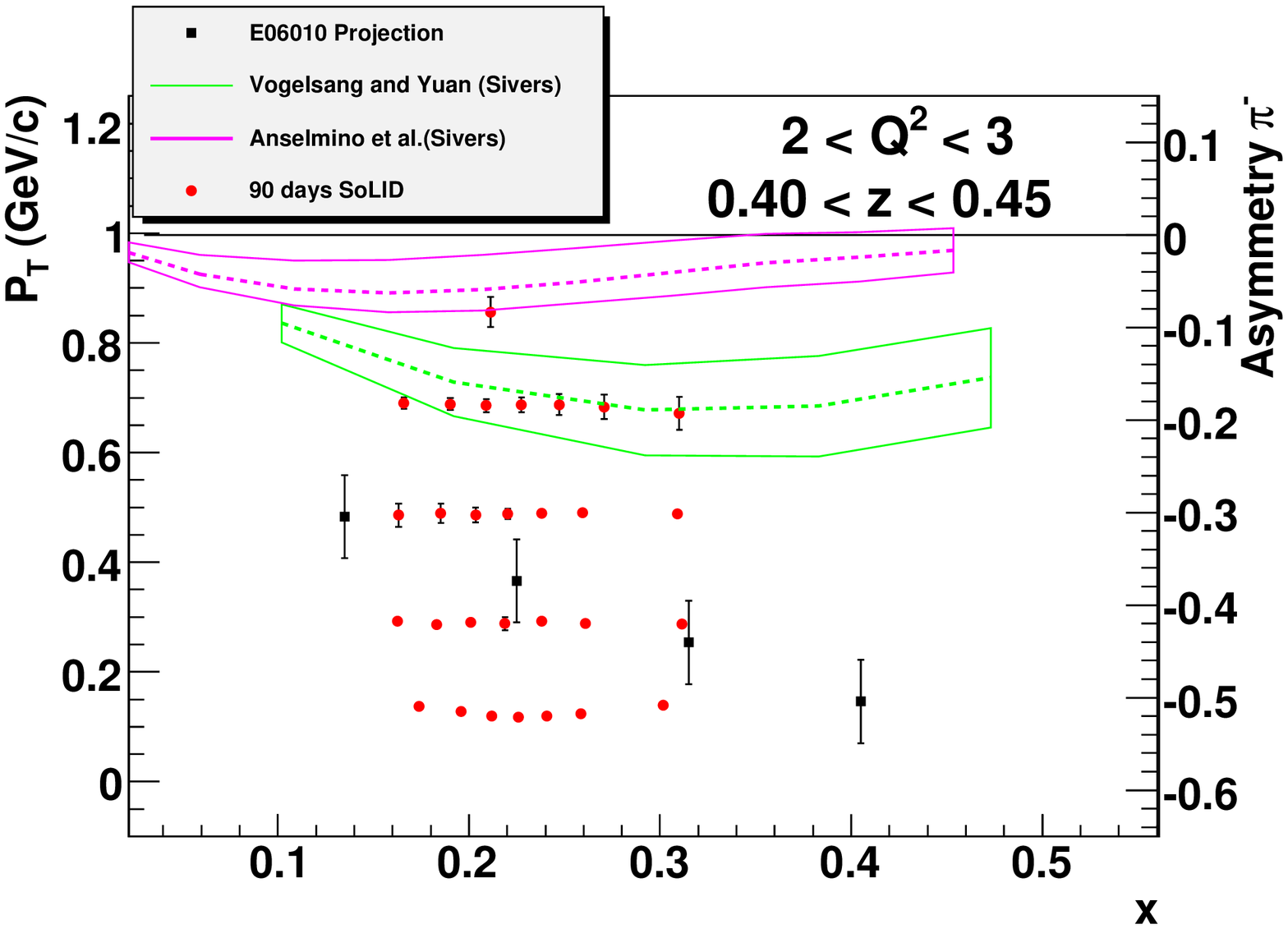}} 
\caption{\label{fig:cdf_proj4} 12 GeV Projections with SoLID. $\pi^-$
  Sivers asymmetries at 0.45 $>$ $z$ $>$ 0.4, 3 $>$ $Q^2$ $>$ 2.}
\end{figure}

The projected results for $\pi^+$ ($\pi^-$) Collins and pretzelosity 
asymmetries at one typical kinematic bin, 
0.45 $>$ $z$ $>$ 0.4, 3 $>$ $Q^2$ $>$ 2, are shown in
Fig.~\ref{fig:cdf_proj1}  (Fig.~\ref{fig:cdf_proj2}) together 
with theoretical predictions of Collins asymmetries from Anselmino {\it et
  al.}~\cite{Anselmino:pv}, Vogelsang and Yuan~\cite{Vogelsang:pv} and
predictions of Collins/pretzelosity asymmetries from Ma and collaborators~\cite{Huang:2007qn,Ma:2009bq}, and
Pasquini~\cite{Pasquini:pv,Boffi:2009sh}. The projected E06-010 results are
shown as black points.  The $x$-axis is $x_{bj}$, and the $y$-axis 
on the left side
is $P_T$, the transverse momentum of the hadron. 
The y-position of the projections
shows the average $P_T$ value for the corresponding kinematic
bin. The $y$-axis on the right side
shows the scale of the asymmetry.
The statistical uncertainties follow the scale on the right side
of the y-axis as well as the theoretical calculations. 
The corresponding Sivers asymmetries for $\pi^+$ ($\pi^-$) 
are shown in Fig.~\ref{fig:cdf_proj3} (Fig.~\ref{fig:cdf_proj4}).

\begin{figure}[htbp] 
\resizebox{150mm}{200mm}{\epsfig{figure=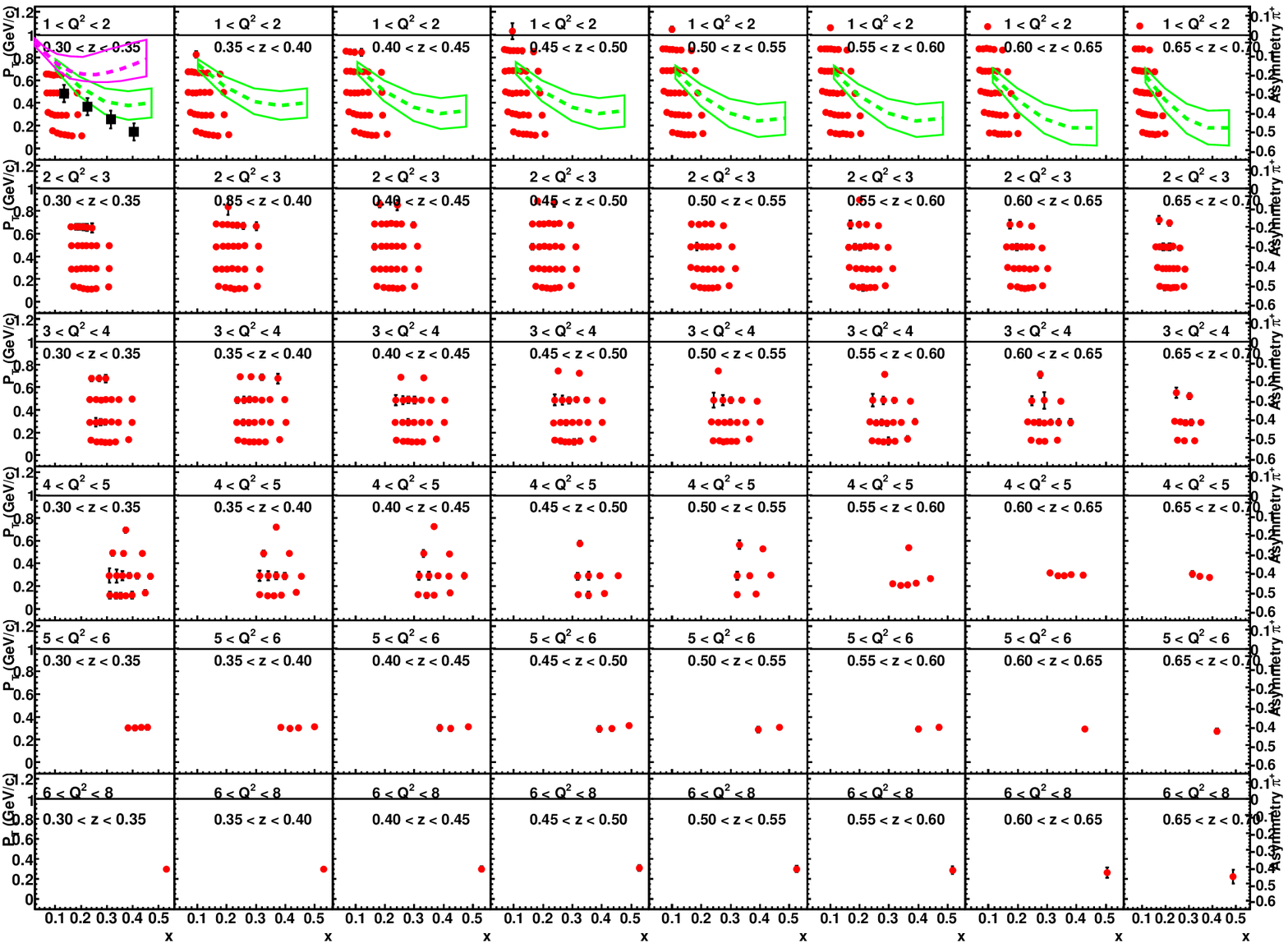,angle=90}} 
\caption{\label{fig:cdf_proj7} 12 GeV Projections with SoLID. $\pi^+$ 
  Sivers asymmetries for all kinematic bin in terms of
  different $z$ and $Q^2$ bin. }
\end{figure}

\begin{figure}[htbp] 
\resizebox{150mm}{200mm}{\epsfig{figure=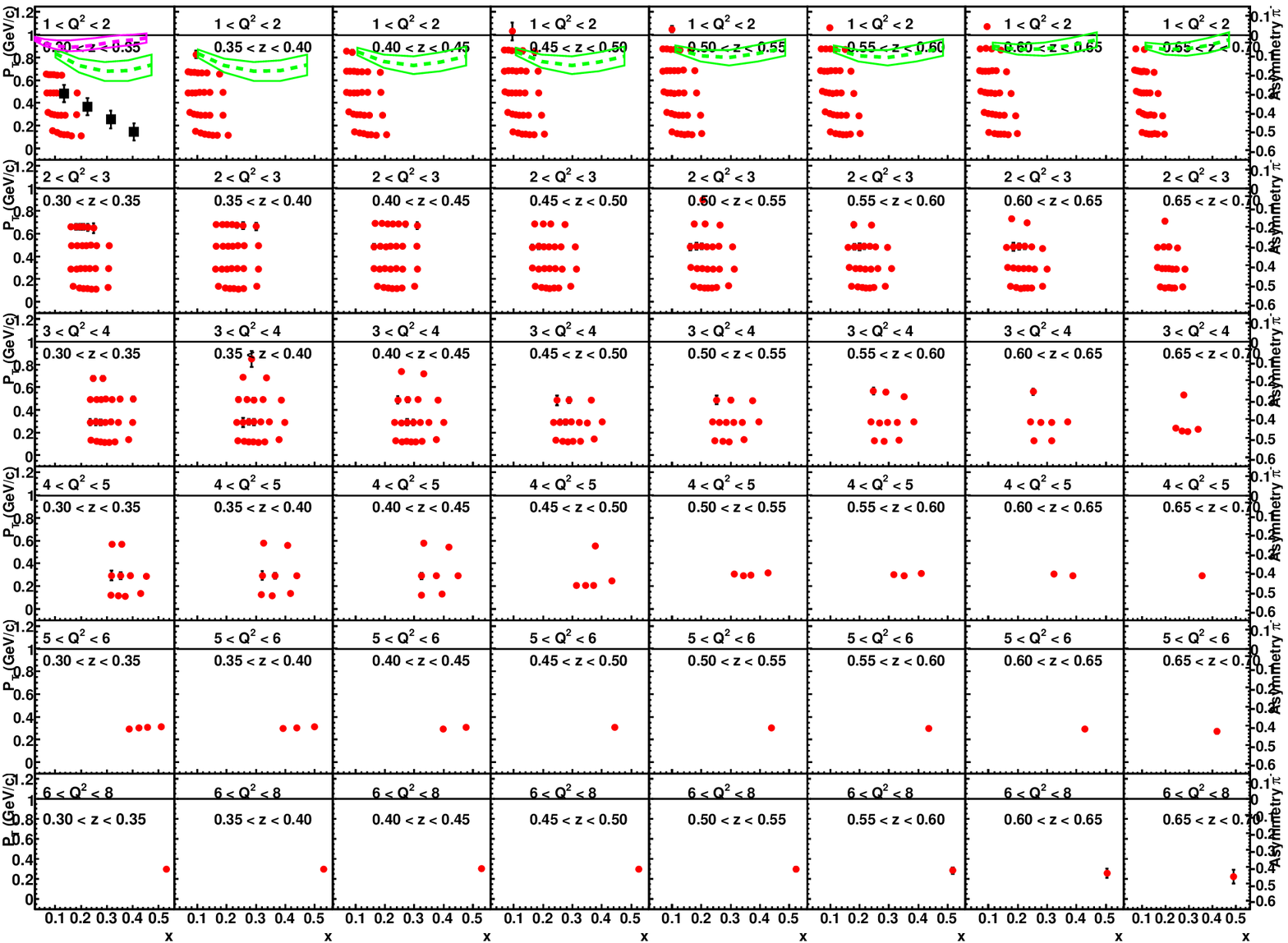,angle=90}} 
\caption{\label{fig:cdf_proj8}12 GeV Projections with SoLID. $\pi^-$ 
  Sivers asymmetries for all kinematic bin in terms of
  different $z$ and $Q^2$ bin.}
\end{figure}

  The complete projections for $\pi^+$ ($\pi^-$) Sivers
asymmetries are shown in terms of 4-D ($x$, $z$, $P_T$ and $Q^2$)
kinematic bins in Fig.~\ref{fig:cdf_proj7}
(Fig.~\ref{fig:cdf_proj8}). Theoretical predictions of Sivers
asymmetries from Anselmino {\it et
  al.}~\cite{Anselmino:pv} and Vogelsang and
  Yuan~\cite{Vogelsang:pv} are shown in comparison~\footnote{The projections
    on Sivers asymmetry are shown separately from the projections of
    Collins and pretzelosity asymmetries, since the projection
    statistical uncertainties are different due to $\phi_h$
    coverage.}. 
We also include the current E06-010 projections in the first panel.
The complete projections for $\pi^+$ ($\pi^-$) Collins/pretzelosity
asymmetries in terms of 4-D ($x$, $z$, $P_T$ and $Q^2$) are not presented 
here, but the statistics are comparable to those presented for the Sivers case.

By extending the study of SIDIS to a real 4-D manner ($x$, $Q^2$, $z$ and
$P_T$), these new results will advance significantly our 
understanding of transverse spin physics.  
The Sivers distribution and the pretzelosity 
distribution functions, crucial to understand 
relativistic effects and the role of quark orbital angular momentum 
inside the nucleon, will also be mapped precisely in four dimensions in this new experiment. The large $P_T$  
region covered in this new experiment is important in testing 
various theoretical approaches. 
Further, a relatively large $Q^2$ region will allow for a study of the
$Q^2$ evolution and 
higher-twist contributions~\footnote{Higher twist effects can also be
  observed with contributions with $\sin(\phi_S)$ and $\sin(2\phi_h -
  \phi_S)$ angular modulation.}.
Finally, extending the measurement of the transversity
distribution to the large $x$ region is essential to extract
the tensor charge. 
A combined analysis between these proposed new 
neutron results and the future proton
results, will be carried out to extract the tensor charge of 
the u and d quark. The high statistics data from both the proton and the
neutron will lead to a unprecedented precise determination of tensor
charge. 
Such a precision will provide important tests of Lattice QCD predictions 
for the tensor charge.

\section{Summary}

\textcolor{black}{
Probing TMDs, in particular the transversity distribution, the least known 
leading-twist quark distribution function, which is non-zero upon integrating over the quark 
transverse momentum, is among the goals of several ongoing and future
experiments.  The experimental study of TMDs, which is now only at its 
inception, promises to have a very exciting future.  Understanding 
the TMDs is certainly a complex task which demands 
major efforts in different laboratories in studying many different processes ranging over 
a wide kinematic region.}
 This is a fast evolving field with growing interest worldwide. 
The new (SoLID) experiment discussed in this paper will provide SSA data 
with excellent statistical and systematic precisions in 4-D ($x$,$z$,$P_{T}$, and $Q^2$) 
over a large kinematic range. These data will significantly advance our understanding of TMDs and QCD. 
Measurements with SoLID, combined with the
CLAS12 measurements~\cite{CLAS2010LO,CLAS2010AA,CLAS2010BB,CLAS2010CC,CLAS2010DD} using polarized proton and deuteron targets will provide an unprecedented opportunity to obtain a three-dimensional map of the Collins and Sivers asymmetries in the kinematical region 0.1 $<$ $x$ $<$ 0.5, 0.3 $<$ $z$ $<$ 0.7 with $P_{T}$ $<$ 1.5 GeV, necessary to precisely determine the nucleon's partonic substructure.
This new experiment 
will have a major impact on other related programs and  
particularly on the design of future facilities with the study of TMDs
as one of their important physics goals, for example the electron ion collider (EIC), 
the FAIR project at GSI. It
will also help to move theory forward in understanding and in modeling 
the quark TMDs significantly. 

\section{Acknowledgements}
This work is supported in part by 
the U.S. Department of Energy under contracts, DE-AC05-84ER40150,
modification No. M175, under which the Southeastern Universities
Research Association operates the Thomas Jefferson National Accelerator
Facility, DE-FG02-03ER41231 (H.G.), and DE-FG02-07ER41460 (L.G.).

\bibliographystyle{h-physrev3}
\bibliography{solenoid_11gev_epja_LG_update_9_eprint}

\end{document}